\documentclass{aa}
\usepackage{graphicx}
\begin{document}

\title{The T\,Tauri star RXJ\,1608.6-3922 -- not an eclipsing binary
       but a spotted single star 
  \thanks{Based on observations obtained at the European Southern
	Observatory at La Silla, Chile in program 62.I-0418, 63.I-0096,
	64.I-0294, 65.I-0275 and on observations obtained at the MSSSO
	Observatory at Siding Springs, Australia.}}

%   \subtitle{}

\author{V. Joergens \inst{1}
          \and E. Guenther \inst{2}
	  \and R. Neuh\"auser \inst{1}
          \and M. Fern\'andez \inst{3} 
          \and J. Vijapurkar \inst{4}	
          }

   \institute{Max-Planck-Institut f\"ur Extraterrestrische Physik,
              Giessenbachstrasse 1, D-85748 Garching, Germany\\
              email: viki@mpe.mpg.de
         \and
              Th\"uringer Landessternwarte Tautenburg,
              Karl-Schwarzschild-Observatorium,
              Sternwarte 5,
              D-07778 Tautenburg,
              Germany
         \and
              Instituto de Astrof\'{\i}sica de Andaluc\'{\i}a (CSIC),
              Apdo. 3004,
              E-18080 Granada,
              Spain     
         \and
	      Homi Bhabha Centre for Science Education
              (Tata Inst. of Fundamental Research),
              V. N. Purav Marg
              Mankhurd, Mumbai 400088,
              India~  
             }

   \offprints{V. Joergens}

   \date{Received; accepted}

\titlerunning{The T Tauri star RXJ\,1608.6-3922}
\authorrunning{Joergens et al.}

   \abstract{
	High-resolution spectroscopy and photometric monitoring
	of the pre-main sequence star \object{RXJ1608.6-3922} shows
	that it is not an eclipsing binary, as previously claimed.
	Radial velocity measurements covering suitable time spans
	in order to detect a spectroscopic binary with the claimed period
	of about 7 days have been performed.
	The scatter of the radial velocity does not exceed
	2.4\,km\,s$^{-1}$, defining an upper mass limit of 
        24 M$_{\mbox{\tiny Jupiter}}$
	for any eclipsing companion orbiting
        this star with the claimed period. 
	Photometric observations of RXJ\,1608.6-3922 in 7 consecutive nights
        (i.e. as long as the claimed orbital period)
	reveal brightness variations of the order of 0.2\,mag
	with a period of 3.6\,days. The shape of the detected
	light curve differs from a light curve
        of the star recorded in 1996.
        The small variations of the radial velocity,
	the variable shape of the light curve, 
	as well as (B-V) color variations
        suggest
	that the flux of RXJ\,1608.6-3922 is modulated
	by spots on the stellar surface 
	with a rotational period of 3.6\,days.
        The stellar activity of this star seems to be highly variable,
        taking into account
        the variable shape of the light curve, with an amplitude varying 
        from 0.5 to 0.2\,mag in a few years, as well as hints for 
        a variable H$_{\alpha}$ equivalent width.
        }

   \maketitle

%
%________________________________________________________________

\section{Introduction}
Photometric observations 
of the young, pre-main-sequence (PMS) object
\object{RXJ1608.6-3922} revealed brightness variations typical for an 
eclipsing system (Wichmann et al. 1998, hereafter W98). 
The light curve recorded in 1996 in
the V band in nine consecutive nights
displays two deep minima of the order of 0.5\,mag and 0.3\,mag
and the authors found a period of 7.2 days.
Similar amplitudes of a primary and a secondary eclipse
indicate similar luminosities and therefore a mass ratio near unity.
Such systems are very likely double-lined spectroscopic binaries (SB2).

The detection of an eclipsing SB2 pre-main-sequence star would 
be highly interesting since it would allow the direct determination of
the masses of the two stellar components. The mass is the most fundamental 
parameter for the evolution of a star and is therefore a very important
input parameter for evolutionary theories. So far there are only
a few late-type low-mass PMS systems known with accurately determined masses:
an eclipsing SB2 in Orion, RXJ\,0529.4+0041, with stellar 
masses of 1.30\,M$_{\odot}$ and 0.95\,M$_{\odot}$ (Covino et al. 2000); 
a 1.6\,M$_{\odot}$ T~Tauri star in an eclipsing SB2 within the 
triple system TY CrA (Casey et al. 1998);
a 1.1\,M$_{\odot}$ PMS star in the detached eclipsing SB2 EK\,Cep
(Popper 1987);
the eclipsing SB2 RS\,Cha, with masses of 1.86\,M$_{\odot}$ 
and 1.82\,M$_{\odot}$ (Anderson 1991), who was recently classified 
as PMS object by strong x-ray emission (Mamajek et al. 2000);
and astrometric measurements of the PMS spectroscopic 
binary \object{NTT045251+3016} yielded masses of 1.4\,M$_{\odot}$ and 
0.8\,M$_{\odot}$ (Steffen et al. 2000). 

Due to the lack of a large sample of well determined stellar properties
for stars in the PMS phase, the 
large discrepancies between the different sets of current 
PMS evolutionary models are not understood and the tracks could not be 
calibrated yet. 

In order to explore if RXJ\,1608.6-3922 is a spectroscopic
binary and to confirm the eclipses,
we carried out high-resolution spectroscopy and 
photometric monitoring and report here on the results of these observations.  

\section{Data acquisition}

\subsection{Spectroscopy}

Using the Echelle spectrograph FEROS (Fiber fed Extended Range
Spectrograph) on the 1.5\,m telescope at La Silla, we have obtained 10
spectra of \object{RXJ1608.6-3922} in March and June 1999 and March 2000. 
FEROS is optimised for high
precision measurements of radial velocities.
The spectra cover the wavelength
region between about 3600\,{\AA} and 9200\,{\AA} and have a resolution
of $\rm \lambda / \Delta \lambda=48000$. 
Precise radial velocities have been measured
by cross-correlating the spectra of \object{RXJ1608.6-3922} 
with spectra of the radial velocity standard HR\,5777, 
using the spectral range between 4000 and 6720\,\AA. 
The spectra of HR\,5777 have been obtained also with FEROS in the same 
observing run. 
The radial velocity
of this standard is known to be stable and has been determined with an
accuracy of $\rm 55\,ms^{-1}$ (Murdoch \cite{murdoch93}).

\subsection{Photometry}

We monitored RXJ\,1608.6-3922 photometrically in seven consecutive nights
in April 2000 in Chile and Australia in the Johnson V and B filter.
CCD images were obtained in Chile at the Danish 1.5\,m telescope at ESO,
La Silla, between April 21 and 28 with the imaging camera DFOSC.
The observations in Australia were carried out with a CCD camera
at the MSSSO 40\,inch telescope at Siding Springs observatory between 
April 23 and 26.
At MSSSO only V band images have been taken.

We performed aperture photometry for RXJ\,1608.6-3922 and several
reference stars in the field with the qphot package of 
IRAF\footnote{IRAF is distributed by the National Optical
   Astronomy Observatories,
   which is operated by the Association of Universities for Research in
   Astronomy, Inc. (AURA) under cooperative agreement with the National
   Science Foundation.}. 
%An aperture of about 5 times the FWHM has been used. The 
%sky background was determined from a concentric ring outside the aperture 
%and subtracted from the raw magnitude of the star.
Differential photometry has been carried out, 
allowing us to compensate for variable atmospheric conditions and also
to combine the Australian and Chilean data. 
%
%The observations at the Indian VBO telescope did not
%yield useful data due to problems with scheduling,
%weather and techniques.

\section{Results}

\subsection{Radial velocities}
\begin{figure}[t]
\vbox{
 \includegraphics[width=0.49\textwidth]
%{rv_curve_new1.ps}
{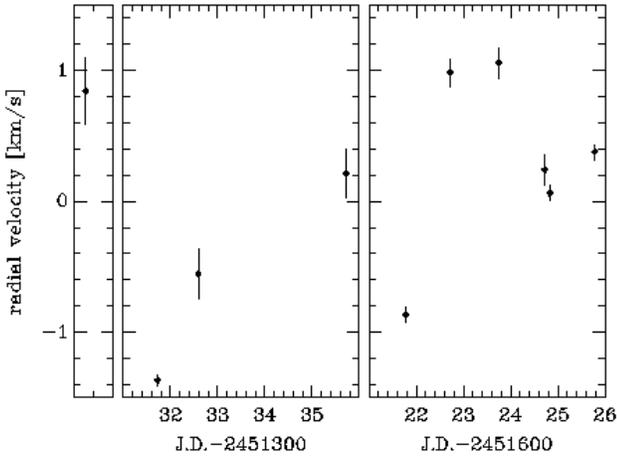}
\hfill
}
\caption{\label{rvs} Radial velocities of RXJ1608.6-3922 determined
from high-resolution spectra taken with FEROS 
between March 1999 and March 2000.
The scatter of the radial velocity is smaller than 2.4\,km\,s$^{-1}$.
The observations in 5 \emph {consecutive} nights (right panel)
cover more than half of the claimed period of 7.2\,d. Therefore
these radial velocity measurements exclude a 7.2\,d orbit with a higher radial velocity amplitude than 2.4\,km\,s$^{-1}$.
The unlabeled data point in the left panel
corresponds to a JD of 245 1264.3.
}
\end{figure}

The main result of our spectroscopic 
observations is that we detected almost no variations of the radial velocity 
at all (see Fig.~\ref{rvs}). 
The scatter does not exceed 2.4\,km\,s$^{-1}$ for 10 spectra 
obtained within one year.
In particular this is the case for 
radial velocities measured in 5 \emph{consecutive} nights 
(right panel, Fig.~\ref{rvs}) covering more than half of the claimed period 
of 7.2\,days. Hence, the semiamplitude of 
the detected variations cannot be larger than 2.4\,km\,s$^{-1}$ for a
7.2\,d orbit.

This result comes as a surprise for a system showing (apparently)
a primary and secondary eclipse of 0.5\,mag and 0.3\,mag, respectively
(W98). On the one hand
only a very low-mass companion causes such small radial velocity variations,
whereas on the other hand only a relatively luminous companion 
causes primary and secondary eclipses of almost similar amplitudes.
Therefore it is unlikely that \object{RXJ1608.6-3922} is an eclipsing
spectroscopic binary.

However, the detected radial velocity 
variations are significant 
but can easily be explained by stellar activity. 

Based on the FEROS spectra, we determined the projected rotational velocity.
Using the telluric lines for determining the instrumental profile of
the spectrograph and assuming a solar-like center-to-limb variation,
we derived a v\,sini of 22.4\,$\pm$\,1.9 km\,s$^{-1}$.
This is in 
agreement with the values of 21.8\,km\,s$^{-1}$ and 20\,km\,s$^{-1}$,
estimated by Wichmann et al. (1999) 
by cross correlation and Fast Fourier Transformation, resp., 
using a template spectrum. 
Furthermore the radial velocity determined by these authors 
of 1.2\,km\,s$^{-1}$ agrees well with our values 
(cp. Fig.~\ref{rvs}).

\subsection{The primary mass and the mass limit for a companion}

From the radial velocity variations an upper mass limit for an eclipsing
companion can be derived.
Assuming a mass of M$_1$=0.78\,M$_{\odot}$ for the primary 
(Wichmann et al. 1997, hereafter W97) and an orbital period of 7.2\,d, 
a radial velocity semiamplitude of 2.4\,km\,s$^{-1}$ defines 
an upper mass limit of 20\,M$_{\mbox{\tiny Jup}}$ for any eclipsing
companion to this star. 
The adopted primary mass was estimated by W97
from the star's bolometric luminosity and
effective temperature and comparison with evolutionary tracks 
of D'Antona\,\&\,Mazzitelli (1994) using Canuto and Mazzitelli convection 
and Alexander opacities.

If there is a companion and it contributes significantly to the 
total luminosity of the system, the  
luminosity of the primary L$_1$ would be considerably smaller
than the estimated luminosity of the unresolved binary.
Considering the worst case, i.e. a secondary which is 
as luminous as the primary (L$_1$=L$_2$=L$_{est}$/2)
and assuming 
similar spectral types, a comparison with the same evolutionary tracks,
as used by W97,
yields M$_1 \approx$ 0.85\,M$_{\odot}$ and a
correspondingly higher mass limit for a companion.

%But talking about uncertainties in the primary mass of 
%\object{RXJ1608.6-3922}, there must be mentioned two other crucial 
%sources of error:
%
%First of all
However,
masses derived by means of comparison with current
PMS evolutionary tracks are highly uncertain 
due to large discrepancies between
different sets of theoretical PMS models and due to a lack of 
observational constraints.
But even with a given evolutionary model, 
the error in the effective temperature of $\sim$\,200\,K
hamper an accurate placement into the Hertzsprung-Russell-diagram (HRD) 
and therefore an accurate
mass determination. Taking the error in T$_{\mbox{\tiny eff}}$ into account
gives a primary mass in the range of 0.6 to 1\,M$_{\odot}$
in the case of a neglectable secondary luminosity and 
0.7 to 0.9\,M$_{\odot}$
if the secondary is as luminous as the primary 
(L$_1=$L$_2$=L$_{est}$/2).
The error in the derived mass is smaller in the latter case because
a smaller luminosity shifts the position of the object in the HRD 
down to the radiative tracks, where the dependence on T$_{\mbox{\tiny eff}}$ is small. 

% -------------------- RV plot -------------------------------
%______________________________________________________________

To summarize, this means the primary mass M$_1$ can have 
values in the range of 0.6 to 1\,M$_{\odot}$, 
considering binarity and errors in T$_{\mbox{\tiny eff}}$,
but not the errors in the adopted evolutionary tracks themselves.
An eclipsing companion pulling on a 0.6\,M$_{\odot}$ primary star 
and causing its radial velocity to change as detected cannot be more 
massive than 17\,M$_{\mbox{\tiny Jup}}$. 
An eclipsing companion to a 1\,M$_{\odot}$ primary star causing 
the detected
radial velocity scatter can have 24\,M$_{\mbox{\tiny Jup}}$ at the most.
The absolute uncertainties in the
mass determination of the primary star due to 
the use of different PMS evolutionary tracks should
not be larger than a factor of two so that the possibly
occulting object would still be substellar.
In any case our spectroscopic observations allow only a very low-mass
substellar companion.

\subsection{Light curves and color variations of \object{RXJ1608.6-3922}}

\begin{figure}[h]
\vbox{
 \includegraphics[height=0.49\textwidth,angle=90]
%{rxj1608V_nice.ps}
{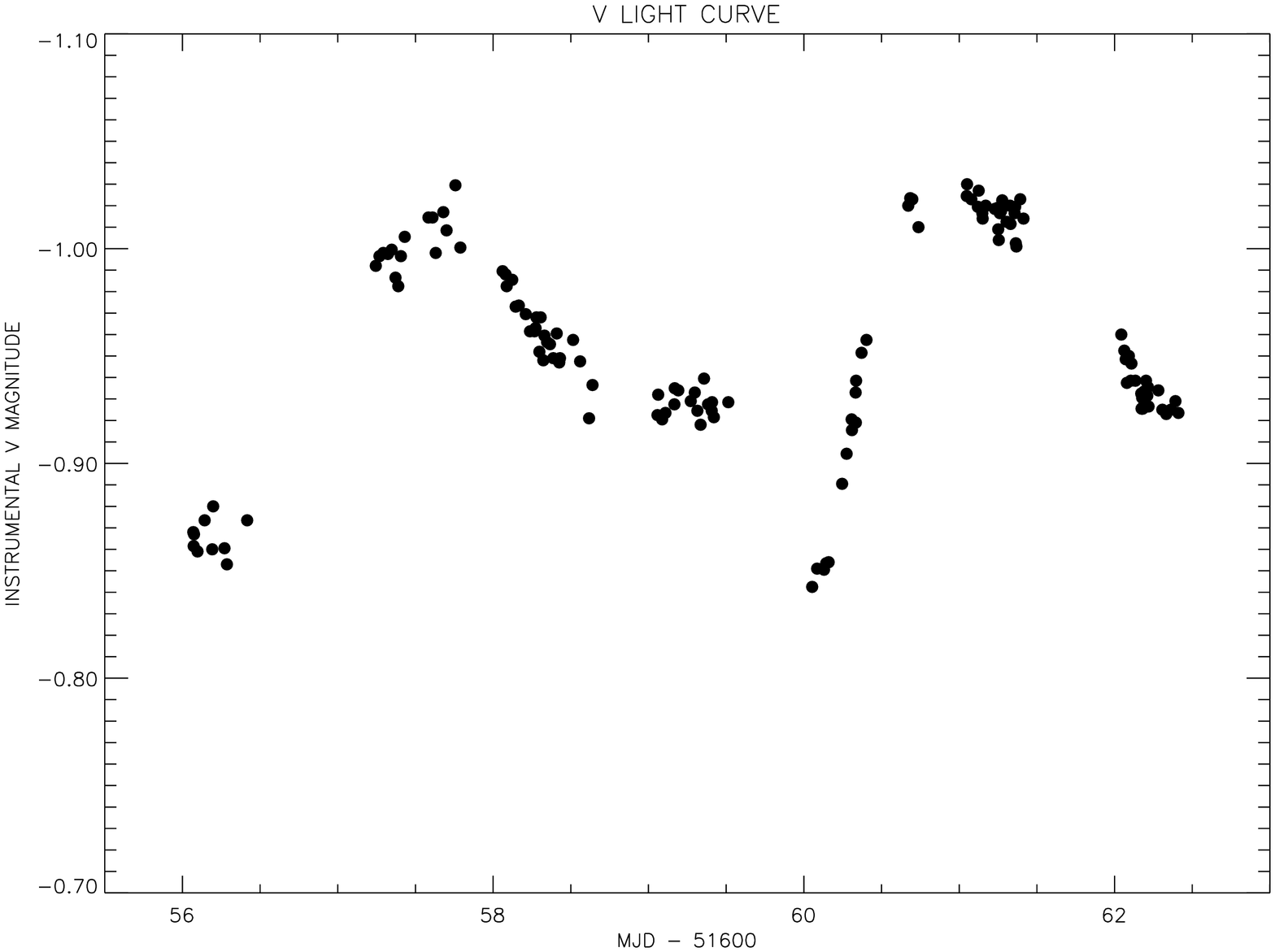}
\hfill
 \includegraphics[height=0.49\textwidth,angle=90]
%{rxj1608B_nice.ps}
{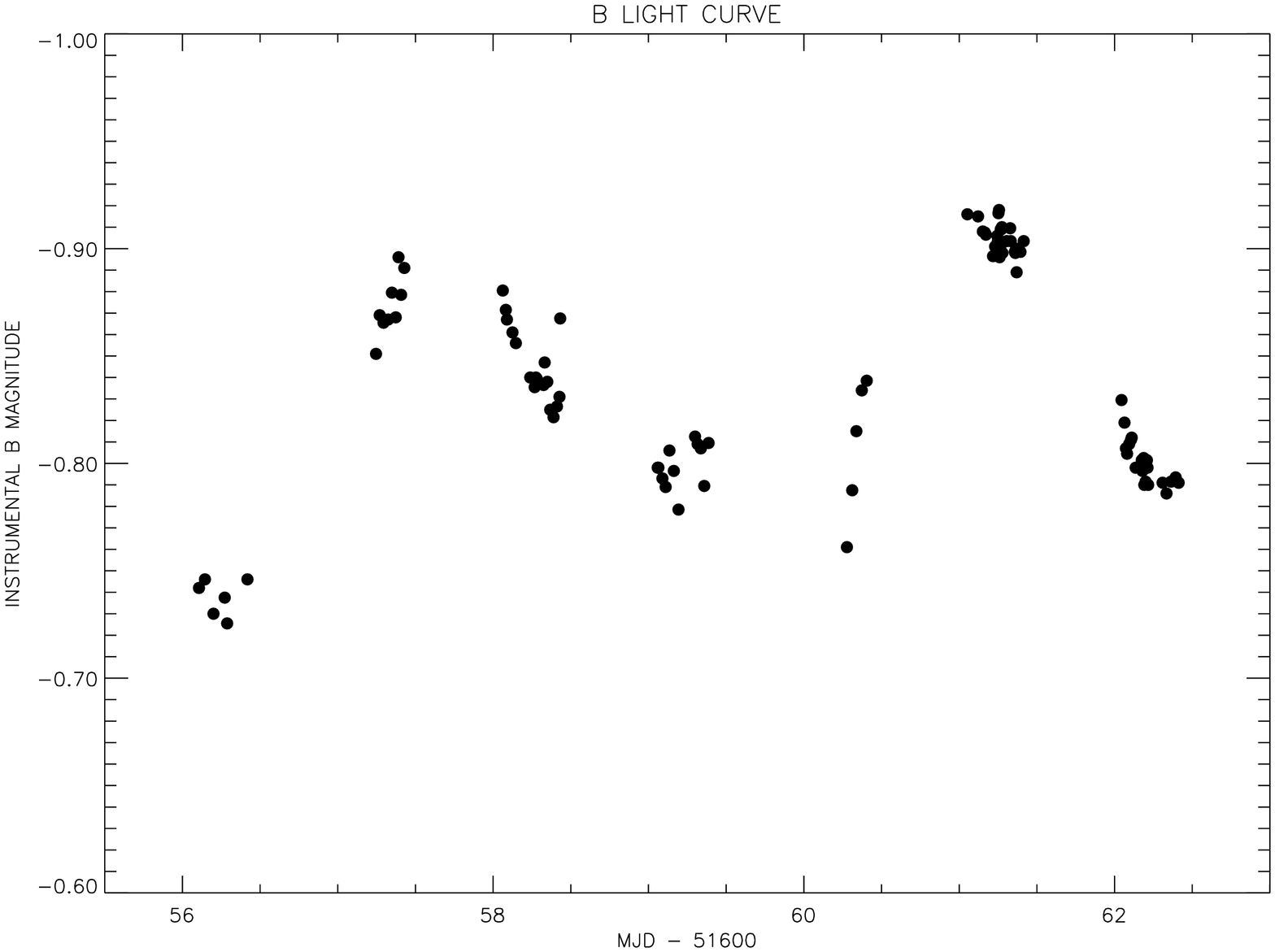}
\hfill
\includegraphics[height=0.49\textwidth,width=5cm,angle=90]
%{rxj1608B_V_nice.ps}
{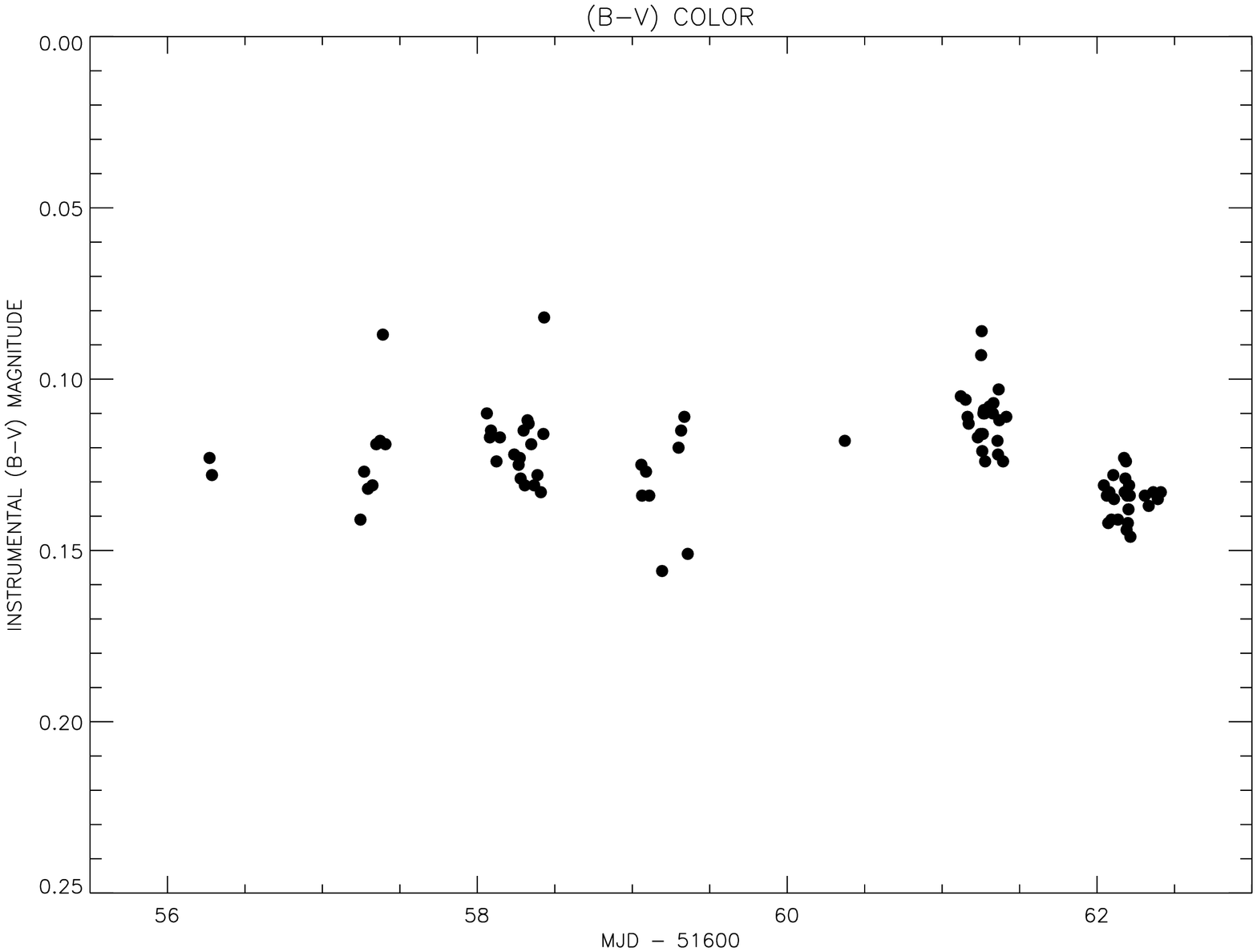}
}
\caption{\label{lcs} Instrumental V and B magnitudes (top and middle)
and instrumental (B-V) colors (bottom) of \object{RXJ1608.6-3922}
obtained at La Silla and Siding Springs. The brightness is increasing
bottom-up.
Both light curves show significant brightness variations. The amplitudes
of the variations are higher at smaller wavelength (i.e. in the B band).
At Siding Springs only V band measurements have been done, leading to 
larger gaps in the B band light curve due to day time observing breaks. 
The (B-V) values show small but significant color variations,
indicating that the system is redder when darker.
} 
\end{figure}

We detected significant brightness variations of about 0.15\,mag in the 
V band and 0.2\,mag in the B band (see Fig.~\ref{lcs}, top and middle). 
The obtained V light curve does not reproduce 
the previously taken V light curve by W98, 
specially we did not detect the deep minima showing a decrease of 
brightness of 0.3 and 0.5\,mag.
Therefore it is dubious whether the 
brightness variations recorded in 1996 by W98 can be explained by 
eclipses of a binary.

% -------------------- light curves  ----------------------------
% ---------------------------------------------------------------

Periodic light variations observed from many T Tauri stars are 
interpreted as modulation of the total flux by spots on the stellar 
surface (see for example Bouvier et al. 1993, COYOTES I). 
The measured photometrical period is a direct tracer of the rotational period
of the star.
Such spot-driven photometric variations come along with
color changes because of the different temperature and therefore
spectral distribution of the spot regions compared to the surrounding 
photosphere.

For \object{RXJ1608.6-3922} we measure larger
amplitudes of the brightness variations in 
the B band than in the V band, i.e. the emission of
the system is shifted towards longer wavelength (= becoming redder)
when the brightness decreases.

This observed
behavior of the color is characteristic for cool as well as for hot
spots. 
It can even be produced by the primary eclipse of a binary 
containing a relatively very low-mass companion, 
whereas the secondary eclipse of such a system
will cause the opposite effect, that is becoming bluer when darker.
A very clear reddening during primary eclipse and blueing during 
secondary eclipse was observed for TZ\,Mensae, 
a binary whose components have masses of 2.5\,M$_{\odot}$ and 
1.5\,M$_{\odot}$ (Andersen et al. 1987).

The brightness of a 
single star with a cool spot on its surface is dimmed when the 
cool spot becomes visible and since it is cool and emitting more 
light at longer wavelength than at shorter the dimming is accompanied
by reddening of the system.

A star with a hot spot shows minimum light when the hot spot is  
\emph {not} visible and since the hot spot contributes more in the 
blue part of the spectral range, its disappearance is accompanied by
relative reddening of the star-spot-system.

Eclipses in a system with a cool, low-mass companion to a hotter,
more massive star can cause color variations along the cycle.
During the primary eclipse, when the low-mass companion  
obscures (part of) the primary, the system gets darker and redder
since (part of) the emission at shorter wavelength from the primary is missing.

During secondary eclipse, when the primary obscures the redder, low-mass 
companion, the system gets darker mainly in the red part of the
spectral range and is therefore bluer in total during this phase.  

In the lower panel of Fig.~\ref{lcs} (B-V) values 
of \object{RXJ1608.6-3922} are shown.
The (B-V) color decreases when the star gets 
darker in V and B, i.e. the star is redder when darker. 
This could be evidence for (cool or hot) spots or primary 
eclipses of a binary with a red, low-mass secondary.
Based on 
the detected color variations we
can exclude that the variations are due to
a secondary eclipse.

We checked the photometric measurements taken in 1996 
in the V filter (W98) and the B filter
(Wichmann, pers. comm.)
on color variations. Both minima show larger amplitudes in 
the B band than in the V band, therefore none of them can 
be explained by the secondary eclipse of a low-mass companion.
However, 
a companion with a higher mass than 24\,M$_{\mbox{\tiny Jup}}$ 
is excluded by our spectroscopy (cp. Sec.\,(3.1)).

% ------------------- period search plot -------------------------
\begin{figure}[h]
\vbox{
 \includegraphics[height=0.49\textwidth,angle=90]
%{v_all_str_period.ps}
{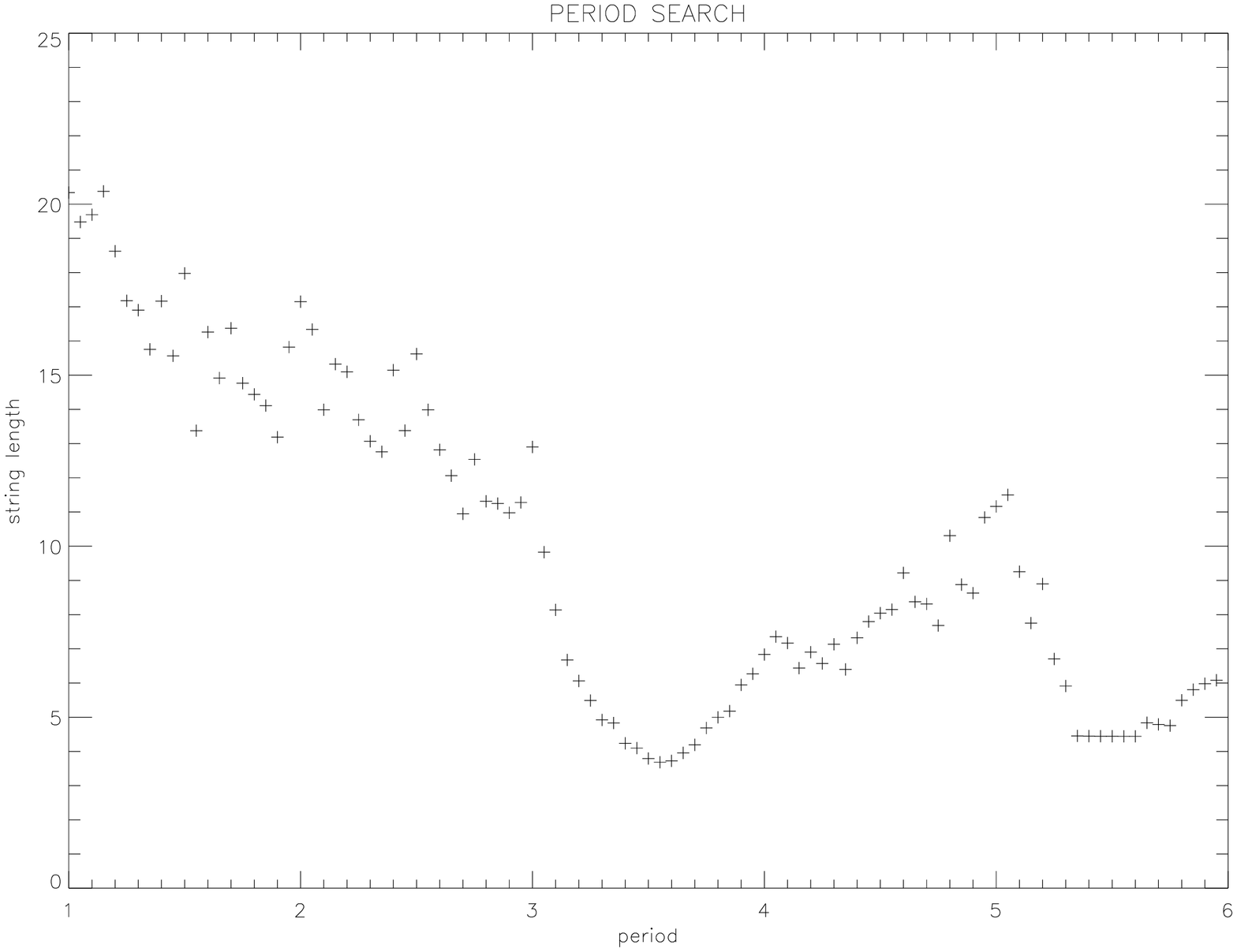}
}
\caption{\label{psearch} Result of a search for periodicity 
in the V photometry using the string-length method (Dworetsky 1983).
Plotted are string-length values calculated for different periods
in the interval from 1 to 6\,days. Clearly visible is the minimum at 
3.55\,days. A second minimum is present at about 5.4\,d but further
analyses show that it is an alias. 
} 
\end{figure}
% ----------------------------------------------------------------

% ------------------- phasefolded lc -----------------------------
\begin{figure}[h]
\vbox{
 \includegraphics[height=0.49\textwidth,angle=90]
%{rxj1608V_ph355_nice.ps}
{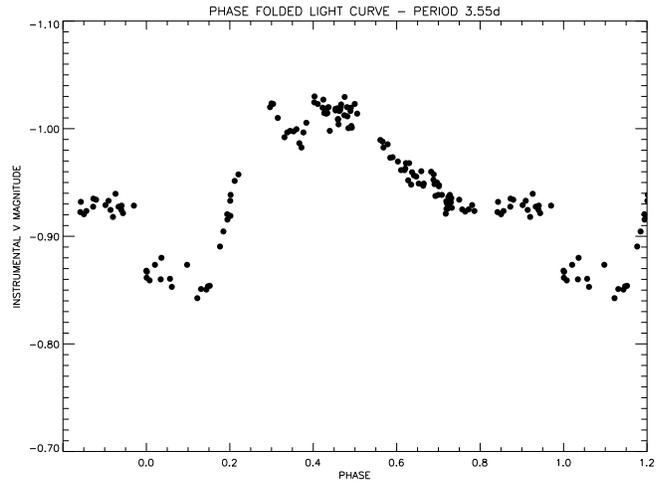}
}
\caption{\label{phlc} 
Phase folded V light curve
of \object{RXJ1608.6-3922} for a period of 3.55 days. 
For clarity 20\% more than one orbit has been plotted
at the beginning as well as at the end of the light curve.
} 
\end{figure}
% ----------------------------------------------------------------
\begin{figure}[h]
\vbox{
 \includegraphics[height=0.49\textwidth,width=5cm,angle=90]
%{rxj1608V_ph720_nice.ps}
{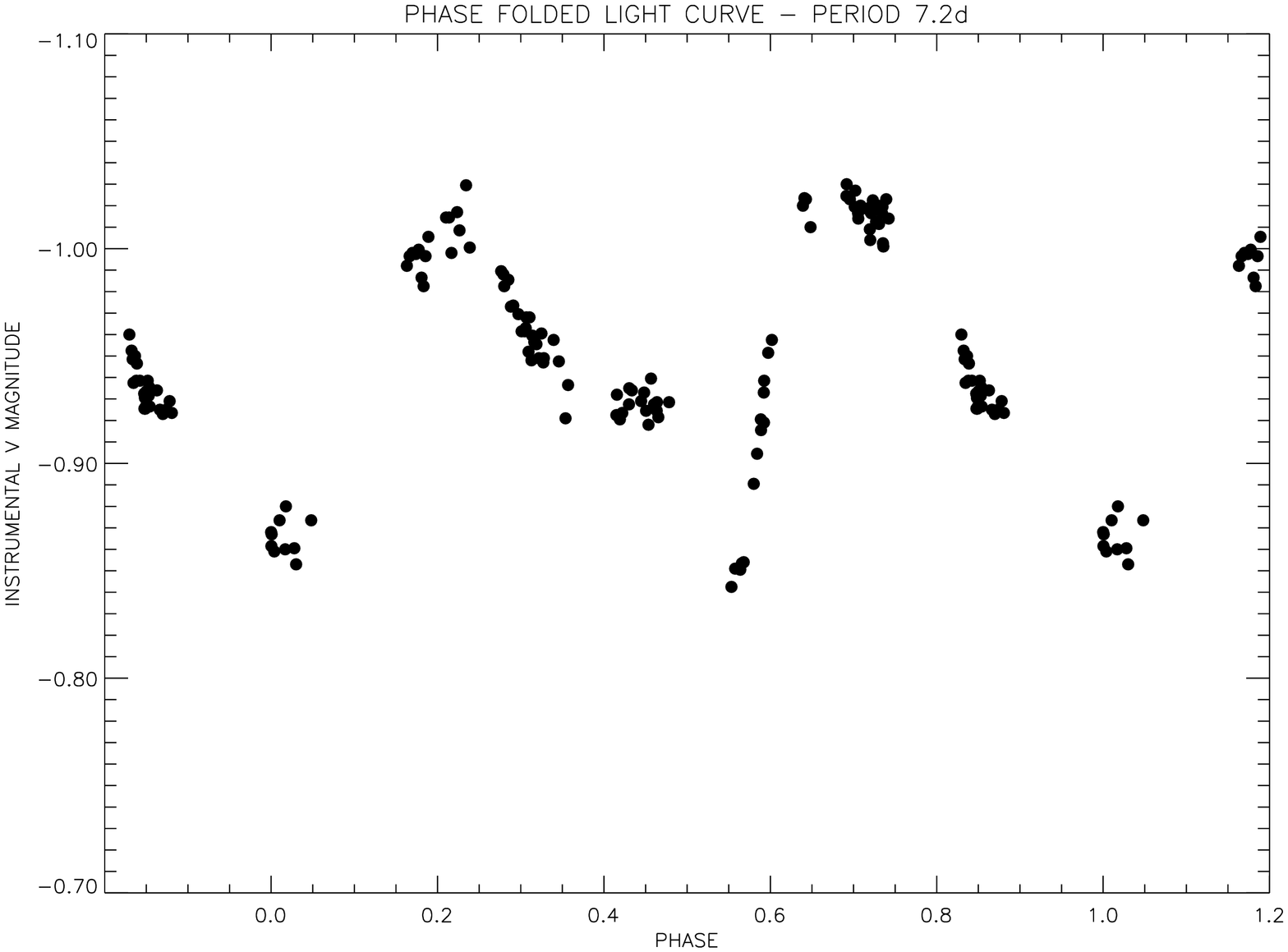}
 \includegraphics[height=0.49\textwidth,width=5cm,angle=90]
%{rxj1608V_ph550_nice.ps}
{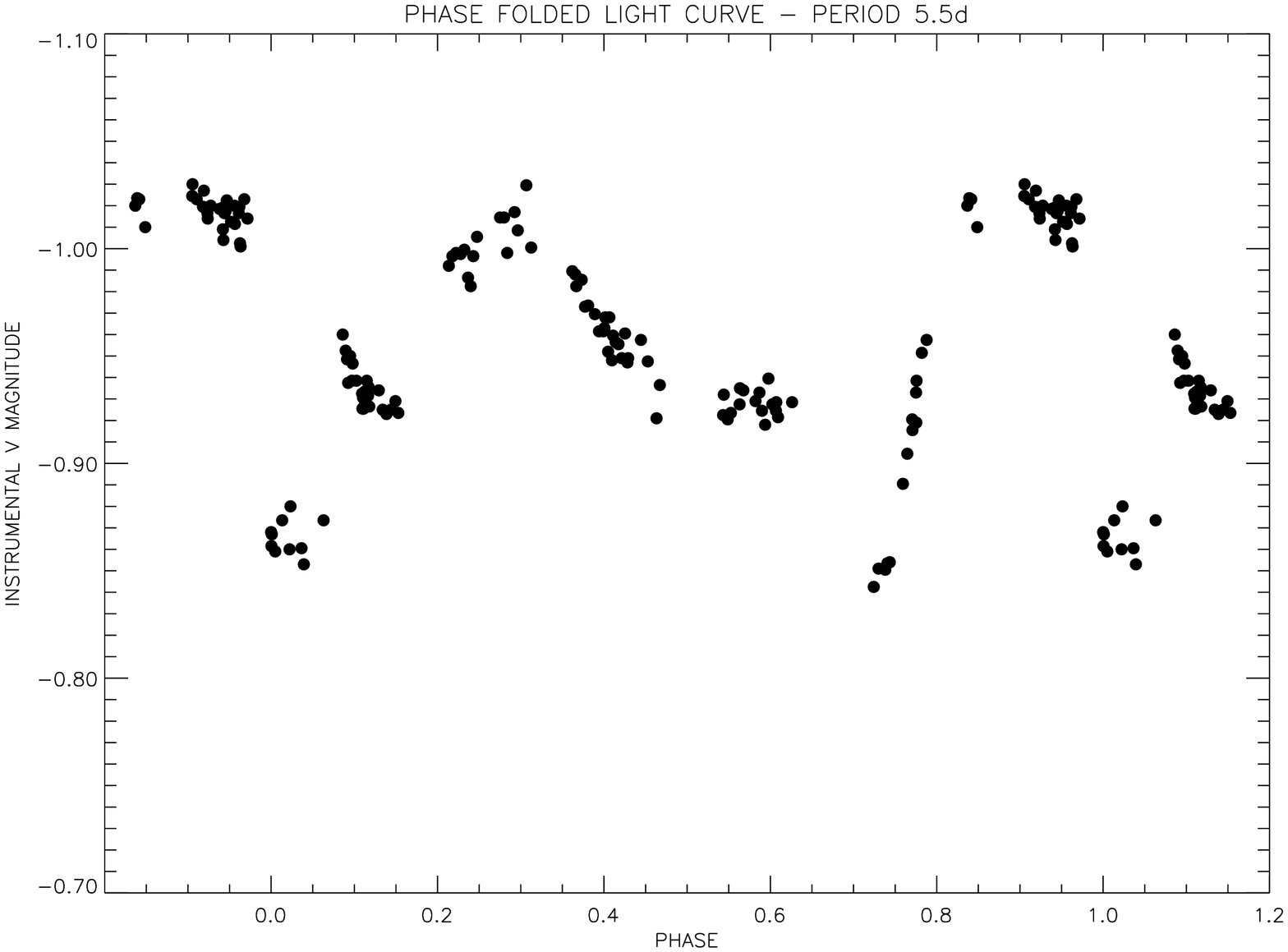}
}
\caption{\label{mehrphlcs}
Phase folded V light curves of \object{RXJ1608.6-3922}.
In the upper panel the light curve is 
folded with a period of 7.2\,days. This is the period claimed by W98.
In the lower panel the light curve is folded with
a period of 5.4\,days. It clearly displays that the second, less
significant period, found
by means of the string-length method is an alias.
} 
\end{figure}
% ----------------------------------------------------------------

\subsection{Period}

We used the string-length method (Dworetsky 1983) in order to 
search for periodicity in our data.
Phase folding the data with a trial period and minimizing
the string-length between successive data points yields a minimum for
the \emph{true} period.
Fig.~\ref{psearch} displays string-length values for periods between 
1 and 6 days. A clear minimum is visible at 3.55\,d and a second, 
less significant minimum at about 5.4\,d.

The light curve of \object{RXJ\,1608.6-3922} 
folded in phase with a 3.55\,d period 
is displayed in Fig.~\ref{phlc}.
It looks quite smooth. 
Phase folding the data with the less significant period of 5.4\,d
(cp. Fig.~\ref{mehrphlcs}, bottom) makes clear that it is an alias.

Our photometric observations cover almost 7 days and
we cannot search for periods longer than the total time basis,
in particular not for the claimed period of 7.2\,d.
However, phasing our V magnitudes with 7.2\,d
shows that our data are also consistent with this period.

\begin{figure}[h]
\vbox{
 \includegraphics[height=0.49\textwidth,angle=90]
%{wichm_ph355.ps}
{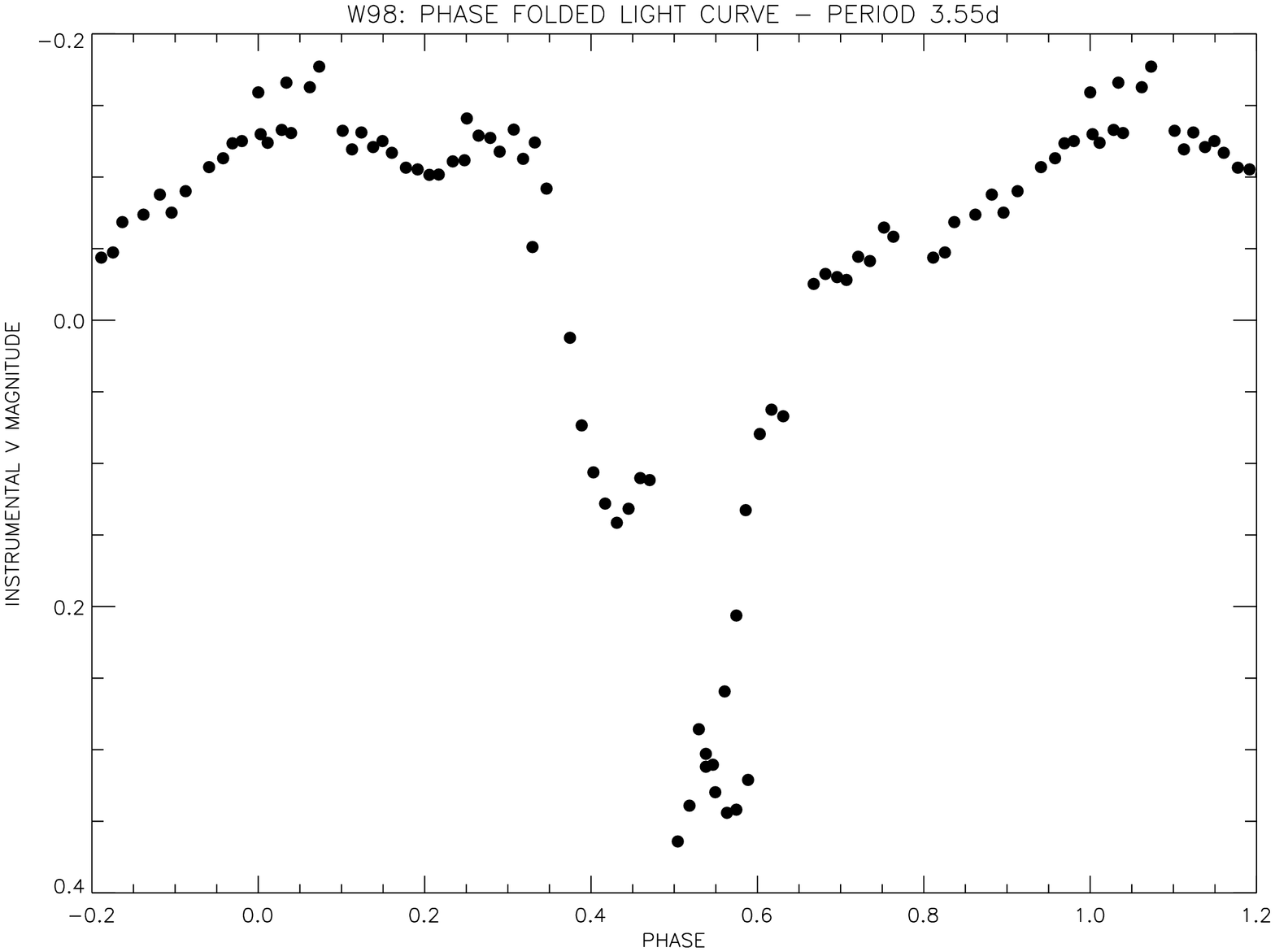}
 \includegraphics[height=0.49\textwidth,angle=90]
%{wichm_ph720.ps}
{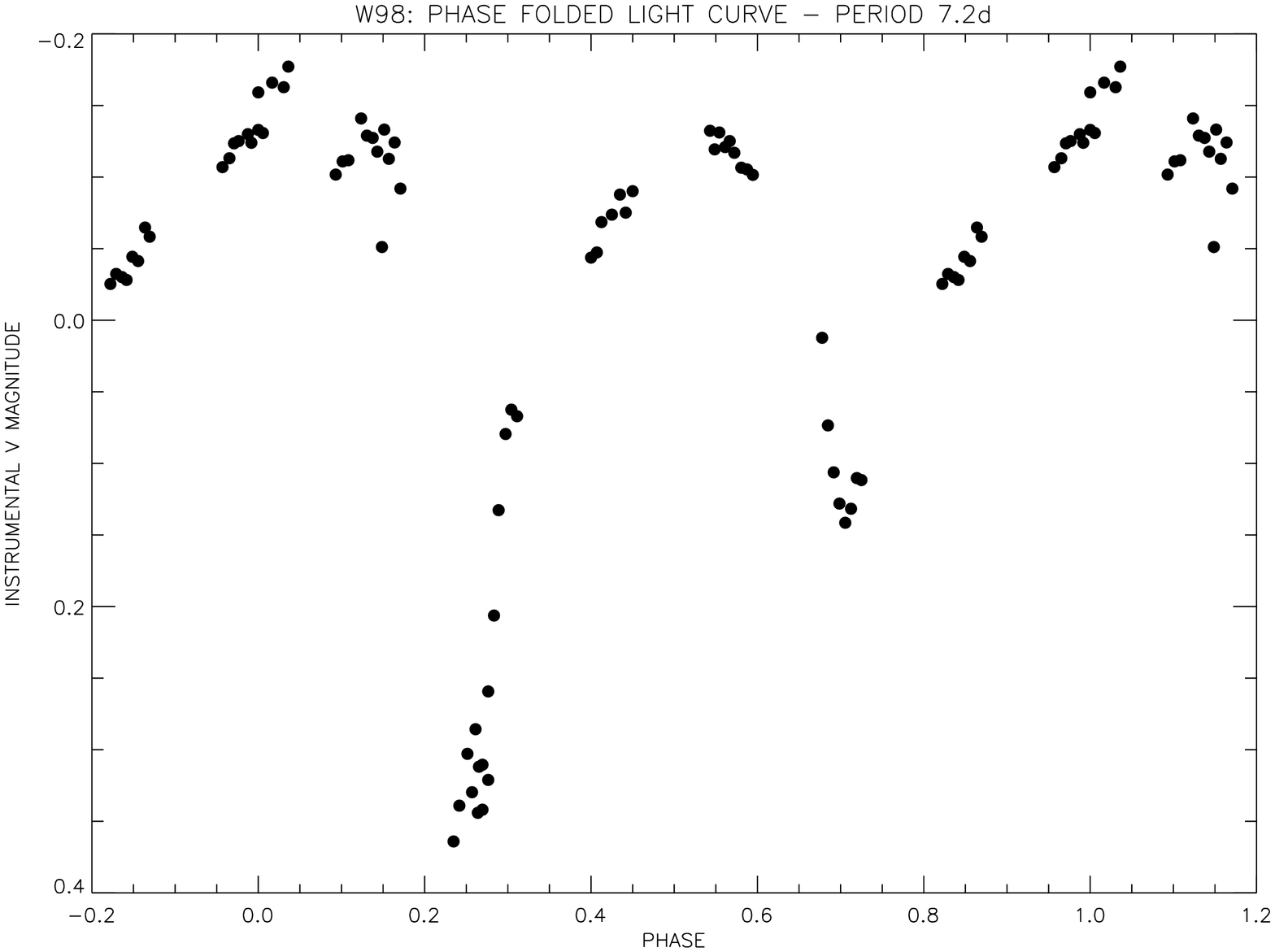}
}
\caption{\label{wichm} 
Displayed are the photometric V band data taken by W98
phase folded with a period of 3.55\,d (top) and 7.2\,d (bottom).
} 
\end{figure}

We phase folded the data taken by W98 with a period of 
3.55\,d and 7.2\,d in order to compare them with our results. 
The folded W98 data are displayed in Fig.~\ref{wichm}. 
Although the 7.2\,d period looks more convincing,
a 3.55\,d period is still in agreement with the data set. 

This means the photometric data of \object{RXJ\,1608.6-3922}
from 1996 as well as from 2000 show brightness variations consistent 
with a 3.55\,d as well as a 7.2\,d period. If these flux modulations
are caused by stellar spots the star is rotating with 
period of 3.55\,d or 7.2\,d.

% ----------------------------------------------------------------
% ----------------------------------------------------------------

\subsection{Constraining the rotational period by means of the stellar 
radius and v\,sin\,i}

An estimate of the rotational period for RXJ1608.6-3922 
based on the radius and the rotational velocity
can help to distinguish between the two found periods. 
Stellar properties for RXJ1608.6-3922 have been published
by W97:
They determined an effective temperature of T$_{\mbox{\tiny eff}}$=4255\,K 
and a luminosity of log(L$_*$/L$_{\odot})$=-0.25
for a distance of 140\,pc. 
From these properties a radius of R$_*$=1.38\,R$_{\odot}$ was derived. 
By comparison with evolutionary tracks from D'Antona \& Mazzitelli (1994)
a mass of M$_*$=0.78\,M$_{\odot}$ was estimated.

% ----------------------------------------------------------------
\begin{table}
\begin{center}
\begin{tabular}{|l|c|c|}
\hline
property  & value      & error \\
\hline
SpT       &   K6       & 0.8 (1.6) subclasses\\
V         & 13.48\,mag & 0.01\,mag \\
B-V       &  1.52\,mag & 0.02\,mag \\        
U-B       &  1.01\,mag & 0.02\,mag \\
V-R$_{\mbox{\tiny C}}$   &  1.01\,mag & 0.02\,mag \\
V-I$_{\mbox{\tiny C}}$   &  1.02\,mag & 0.02\,mag \\
T$_{\mbox{\tiny eff}}$ & 4255.0\,K  & $\Delta\,\log(\mbox{T}_{\mbox{\tiny eff}})=0.02$ \\
%                 &            & $\Delta$ T$_{\mbox{\tiny eff}}$=196\,K                   \\
log(L$_*$/L$_{\odot}$) & -0.25 & $\Delta$\,log(L$_*$/L$_{\odot}$) $\approx$ 0.1\\
A$_{\mbox{\tiny V}}$     &  1.75\,mag & -- \\
R$_*$/R$_{\odot}$ & 1.38 & $\Delta$\,log(R$_*$/R$_{\odot}$) $\approx$ 0.2 \\
M$_*$/M$_{\odot}$ & 0.78 & -- \\
log(age) & 6.47\,yr & -- \\
(v\,$\sin{\mbox{i}})$ & 22.4\,km\,s$^{-1}$ & 1.9\,km\,s$^{-1}$\\
d$_{\mbox{\tiny Hughes93}}$   & 140\,pc     & 20\,pc \\
d$_{\mbox{\tiny W98b}}$  & 190\,pc     & 27\,pc \\
%d$_{\mbox{\tiny wTTS}}$ & 135-165\,pc & \\ 
d$_{\mbox{\tiny K\&H}}$       & 100\,pc     &  -- \\
\hline
\end{tabular}
\caption{\label{stelprop} \small {Stellar properties of RXJ1608.6-3922.
They are published by W97, except the following: spectral type (SpT) 
from Krautter et al. (1997),
v\,sin\,i this paper,
distances of the Lupus star forming region: d$_{\mbox{\tiny K\&H}}$ by 
Knude \& H{\o}g (1998), 
d$_{\mbox{\tiny Hughes93}}$ by Hughes et al. (1993), 
d$_{\mbox{\tiny W98b}}$ by Wichmann et al. (1998b). 
%and d$_{\mbox{\tiny wTTS}}$, mean distance to those ROSAT weak 
%line T Tauri stars with $\Delta$ W$_{Li} > 0$
%in the direction of the Lupus star forming region (Wichmann et al. 1999).
}}
\end{center}
\end{table} 
% ----------------------------------------------------------------

Based on our FEROS spectra, we determined the projected rotational velocity
and found a v\,sini of 22.4\,$\pm$\,1.9\,km\,s$^{-1}$.
With unknown inclination i of the star (we do not assume that 
the system is eclipsing) this value is a lower limit of the rotational
velocity. 

Based on v\,sini and the radius of the star an upper limit of the rotational 
period can be derived:

\begin{equation}
\mbox{P}_{\mbox{\tiny max}}[\mbox{d}]= 50.6145 \frac{\mbox{R} [\mbox{R}_{\odot}]}{(\mbox{v} \sin \mbox{i}) [\mbox{km}\,\mbox{s}^{-1}]}
\end{equation}

%\subsubsection{The radius}
%
Adopting the radius of 1.38\,R$_{\odot}$ estimated by W97
the maximum period 
in the limiting case of an inclination of 90$^{\circ}$
is 3.65\,d.
This excludes definitely a rotational period of 7.2\,d.

W97 estimated the luminosity and therefore the radius for 
a distance of 140$\pm$20\,pc (for Lupus\,1-4, Hughes et al. 1993). 
The Lupus complex consists of five clouds and
\object{RXJ1608.6-3922} is located in Lupus\,3 (for the location of the 
five Lupus clouds compare Tachihara et al. 1996).
There are two other estimations of the distance to the Lupus 
star forming region
based on HIPPARCOS measurements: Wichmann et al. (1998b) 
derive a considerably larger distance of 190\,$\pm$\,27\,pc,
whereas Knude \& H{\o}g (1998) propose that at least Lupus\,2-5 is
as close as 100\,pc.

The luminosity and therefore the radius depends crucially on the 
adopted distance.
A larger distance
gives a larger luminosity L$_*$ and therefore a larger radius R$_*$.
Keeping the rotational velocity constant a larger radius stands for  
a longer period.
Therefore in order to exclude 7.2\,d as possible rotational
period for \object{RXJ1608.6-3922} unambiguously, we must take into account
the largest possible distance.

\subsubsection{Radius estimate from L$_*$ and T$_{\mbox{\tiny eff}}$}

The radius of a star can be calculated from
T$_{\mbox{\tiny eff}}$ and L$_*$ 
\begin{equation}
\mbox{L}_*=4 \pi \mbox{R}_*^2 \sigma \mbox{T}_{\mbox{\tiny eff}}^4   % \Longleftrightarrow  
%      R_*= \sqrt{\frac{L_*}{4 \pi \sigma T_{\mbox{\tiny eff}}^4}}
\end{equation}
and depends on the apparent bolometric
magnitude m$_{bol}$, the distance d and the extinction A$_V$.
%
%\begin{equation}
%M_{bol} - M_{bol,\odot} = -2.5 \log(L_*/L_{\odot}) \quad [mag]
%\end{equation}
%\begin{equation}
%M_{bol} = m_{bol} - 5\log d + 5 -A_V  
%\end{equation}
%\begin{equation}
%m_{bol}=m_V+BC
%\end{equation}
%
%With $M_{bol,\odot}$=4.74\,mag the luminosity can be 
%calculated as follows:
%
\begin{equation}
\log(\mbox{L}_*/\mbox{L}_{\odot})=\frac{(4.74 - \mbox{m}_{\mbox{\tiny V}} +
\mbox{A}_{\mbox{\tiny V}} - \mbox{BC} + 5 \log \mbox{d} -5)} {2.5}
\end{equation}
\begin{equation}
\log(\mbox{L}_*/\mbox{L}_{\odot}) = -4.468 + 2 \log \mbox{d}\,[\mbox{pc}]
\end{equation}

We adopt a bolometric correction of BC=-0.82\,$\pm$\,0.1, as
given by Kenyon \& Hartmann (1995) for a star with spectral type K6. 
The error of BC is about 0.1 due to the given (W97) error  
of $\sim$\,200\,K in T$_{\mbox{\tiny eff}}$.
The error in A$_{\mbox{\tiny V}}$ is about 0.3\,mag
due to errors in the intrinsic color as well as
the applied conversion table.
%and 0.01 for m$_V$. 
% 
%The calculation of the error of equation (6) is:
%
%\begin{equation}
%\Delta\,\log(L_*/L_{\odot})=\frac{\Delta BC}{2.5} + \frac{\Delta m_V}{2.5} + \frac{2 \Delta d}{d \ln10} + \frac{\Delta A_V}{2.5} 
%\end{equation}
%
%\begin{equation}
% \Longleftrightarrow \Delta\,\log(L_*/L_{\odot}) = 0.164 + \frac{2 \Delta d}{d \ln10}  
%\end{equation}
%
%\begin{equation}
%\Delta\,L_*/L_{\odot} = \Delta\,\log(L_*/L_{\odot}) L_*/L_{\odot} \ln 10
%\end{equation}
%
The luminosity is estimated for distances of 100\,$\pm$20\,pc,
140\,$\pm$20\,pc
and 190\,$\pm$27\,pc, respectively:
\begin{equation}
 \begin{array}{rcl}
\bf \log(L_*/L_{\odot})_{100} & = & \bf -0.47 \pm 0.34 \\
\bf \log(L_*/L_{\odot})_{140} & = & \bf -0.18 \pm 0.29 \\
\bf \log(L_*/L_{\odot})_{190} & = & \bf  0.09 \pm 0.29 \\
 \end{array}
\end{equation}
%\begin{equation}
% \begin{array}{rcl}
%\bf (L_*/L_{\odot})_{100}     & = & \bf  0.34 \pm 0.27 \\
%\bf (L_*/L_{\odot})_{140}     & = & \bf  0.66 \pm 0.44 \\
%\bf (L_*/L_{\odot})_{190}     & = & \bf  1.23 \pm 0.82 \\
% \end{array}
%\end{equation}

%
Applying the Stefan-Boltzmann law (2) yields the radius:
\begin{eqnarray*}
\mbox{R}_* [\mbox{R}_{\odot}] = \frac{1}{\mbox{R}_{\odot}[\mbox{m}]} \cdot \sqrt{ \frac{\mbox{L}_*[\mbox{L}_{\odot}] \cdot  \mbox{L}_{\odot}[\mbox{W}]} {4 \pi \sigma[\frac{\mbox{W}}{\mbox{m}^2 \mbox{K}^4}] (\mbox{T}_{\mbox{\tiny eff}}^4 [\mbox{K}^4])} } \\
\mbox{R}_* [\mbox{R}_{\odot}] = 3.3379\cdot10^7 \cdot \sqrt{ \frac{\mbox{L}_*[\mbox{L}_{\odot}]}{\mbox{T}_{\mbox{\tiny eff}}^4 [\mbox{K}^4]} } 
\end{eqnarray*}
%
%\begin{eqnarray*}
%\Delta R_* [R_{\odot}] = \frac{3.3379\cdot10^7}{2} \sqrt{\frac{T_{\mbox{\tiny eff}}^4}{L_*}}\left ( \frac{\Delta L_*}{T_{\mbox{\tiny eff}}^4} - 4 \frac{\Delta T_{\mbox{\tiny eff}} \cdot L_*}{T_{\mbox{\tiny eff}}^5} \right )
%\end{eqnarray*}
%
\begin{eqnarray*}
{\bf R_{*,100}=(1.07 \pm 0.33)\,R_{\odot}} \quad
\mbox{for 100$\pm$20\,pc}   \\
{\bf R_{*,140}=(1.50 \pm 0.36)\,R_{\odot}} \quad
\mbox{for 140$\pm$20\,pc}   \\  
{\bf R_{*,190}=(2.04 \pm 0.49)\,R_{\odot}} \quad
\mbox{for 190$\pm$27\,pc.}
\end{eqnarray*}
The maximum period for these radii
in the limiting case of an inclination of 90\,$^{\circ}$
would be
\begin{eqnarray*}
{\bf P_{max,100}= (2.4\, \pm 0.9)d} \\
{\bf P_{max,140}= (3.4\, \pm 1.1)d} \\
{\bf P_{max,190}= (4.6\, \pm 1.5)d} \\
\end{eqnarray*}
Therefore the largest possible radius would be 2.53\,R$_{\odot}$ 
(for a distance of (190+27)\,pc) and the largest possible upper limit
for the rotational period would be  
P$_{\mbox{\tiny max}}$=6.1\,d, excluding a 7.2\,d period.
\\

\subsubsection{Radius estimate from Barnes-Evans type relation for K giants}

Beuermann et al. (1999, hereafter B99) published recently 
Barnes-Evans type relations
for late-type giants and dwarfs.
Their results for giants are based on angular diameters measured for
27 M-giants and 9 K-giants.
The dwarf relation is based on the measurement of the 
angular diameter for YY\,Gem (eclipsing binary) 
and of estimations of angular diameters  
for 8 M-dwarfs.
The latter are derived by flux fitting of model atmospheres to
observed low-resolution optical/IR spectra (for details see 
B99 and references therein).

The B99 relations are based on V-I$_{\mbox{\tiny C}}$ measurements
and allow us to directly use the V-I$_{\mbox{\tiny C}}$ color of 1.02\,mag 
measured for \object{RXJ1608.6-3922} (W97).
%, without any uncertain 
%transformations between different photometric systems.
The star's luminosity class is
\rm {V} or \rm {IV} (Wichmann, pers. comm.), i.e. being a
dwarf or an intermediate object between a dwarf and a giant.

The radius can be estimated directly from the 
absolute magnitude M$_{\lambda}$ and the visual surface
brightness S$_{\lambda}$:
\begin{equation}
\label{slambda}
\mbox{S}_{\lambda} = \mbox{M}_{\lambda} + 5 \log(\mbox{R}_*/\mbox{R}_{\odot}) 
%\Leftrightarrow  \log(R_*/R_{\odot}) = \frac{S_{\lambda} - M_{\lambda}}{5}
\end{equation}
B99 derived linear relationships {\mbox between} 
S$_{\lambda}$ and V-I$_{\mbox{\tiny C}}$ for M-giants as well as
for M-dwarfs. 
A given relationship for K-giants is based on a 
linear fit to 9 K-giants (V-I$_{\mbox{\tiny C}}<$1.65) tied to the Sun:

\begin{equation}
\label{Kgianteq}
\mbox{S}_{\mbox{\tiny V,Kgiants}}     = 2.86 + 
2.84 (\mbox{V}-\mbox{I}_{\mbox{\tiny C}}) \quad 
\mbox{for V-I}_{\mbox{\tiny C}} < 1.65
\end{equation}
The argument for including the Sun is that Barnes \& Evans (1976) and
Barnes et al. (1977) found
that there is no difference in the surface brightness of giants 
and dwarfs for stars of spectral type 
B-G and that the Sun falls on the giant relation.

The absolute magnitude M$_{\mbox{\tiny V}}$ for RXJ\,1608.6-3922 
is:
\begin{eqnarray*}
%\mbox{M}_{\mbox{\tiny V}}=\mbox{M}_{\mbox{\tiny V}} - 5 \log \mbox{d} + 5 -A_V = 16.73 - 5 \log \mbox{d}\\
%\Delta \mbox{M}_{\mbox{\tiny V}} = \Delta \mbox{M}_{\mbox{\tiny V}} + \Delta A_V + \frac{5 \Delta d}{\mbox{d} \ln 10} \\
\Delta \mbox{M}_{\mbox{\tiny V}} = 0.31 +  \frac{5 \Delta \mbox{d}}{\mbox{d} \ln 10} \\
\mbox{M}_{\mbox{\tiny V}}=6.73 \pm 0.74 \quad \mbox{for d = 100\,pc}   \\
\mbox{M}_{\mbox{\tiny V}}=6.00 \pm 0.62 \quad \mbox{for d = 140\,pc}   \\
\mbox{M}_{\mbox{\tiny V}}=5.34 \pm 0.62 \quad \mbox{for d = 190\,pc} 
\end{eqnarray*}

B99 provides no relation for dwarfs
with (V-I$_{\mbox{\tiny C}}$)\,$<$\,1.65, as is the case for 
\object{RXJ1608.6-3922}.
The giant-relation Eq.\,(\ref{Kgianteq}) may nevertheless be used 
since an object with 
\mbox{(V-I$_{\mbox{\tiny C}}$)=1.02 }
should be well in the regime where the authors expect
the separation between both relations to be very small.
%
%The visual surface brightness estimated with the K-giant relation
%Eq.\,(\ref{Kgianteq}) is S$_{\mbox{\tiny V}}$ = 5.76 and 
%with it and  M$_{\mbox{\tiny V}}$
%the radius becomes:
The radius estimated with Eq.\,(\ref{slambda}) and
(\ref{Kgianteq}) is: 
\begin{eqnarray*}
{\bf R_{*,Kgiants} = (0.64 \pm 0.32)\,R_{\odot}} \quad
\mbox{for 100$\pm$20\,pc}   \\
{\bf R_{*,Kgiants} = (0.89 \pm 0.40)\,R_{\odot}} \quad
\mbox{for 140$\pm$20\,pc} \\
{\bf R_{*,Kgiants} = (1.21 \pm 0.54)\,R_{\odot}} \quad
\mbox{for 190$\pm$27\,pc.}
\end{eqnarray*}
The errors have been estimated by adopting an error of 6\% in 
S$_{\mbox{\tiny V}}$ (cp. B99).
% with the following relation:
%
%\begin{eqnarray*}
%(\Delta R_*/R_{\odot}) = \Delta \log(R_*/R_{\odot}) \cdot (R_*/R_{\odot}) \ln 10 \\
%(\Delta R_*/R_{\odot}) = \frac{1}{5} (\Delta S_V + \Delta \mbox{M}_{\mbox{\tiny V}}) \cdot (R_*/R_{\odot}) \ln 10 \\
%(\Delta R_*/R_{\odot}) = \frac{1}{5} (0.35 + \Delta \mbox{M}_{\mbox{\tiny V}}) \cdot (R_*/R_{\odot}) \ln 10 \\
%\end{eqnarray*}
%
The maximum rotational periods in the limiting case 
of an inclination of 90\,$^{\circ}$
for these radii are:
\begin{eqnarray*}
{\bf P_{max,100}= 1.4\,d \pm 0.8\,d}\\
{\bf P_{max,140}= 2.0\,d \pm 1.1\,d}\\
{\bf P_{max,190}= 2.7\,d \pm 1.4\,d}\\
\end{eqnarray*}
The radii estimated with the B99-K-giant-relation 
are smaller than the radii derived from the luminosity and the 
effective temperature and consequently also the rotational periods.
The largest possible upper limit for the rotational period with
these radius estimates is
P$_{\mbox{\tiny max}}$=4.1\,d, also excluding a 7.2\,d period
and favoring the 3.55\,d period.

\section{Discussion and conclusions}

It was claimed by W98 that the T\,Tauri star \object{RXJ1608.6-3922}
is an eclipsing binary. We reported here on results of high-resolution
spectroscopy and photometric monitoring of this object.

The measured radial velocities of 
\object{RXJ1608.6-3922} are almost constant. Our measurements cover suitable
time spans in order to detect a binary with a 7 day period at the 
maximum separation phases.
The scatter of the radial velocities does not exceed
2.4\,km\,s$^{-1}$, giving evidence that it is not an eclipsing 
binary as claimed previously by W98. 
An eclipsing companion to a 0.6 to 1\,M$_{\odot}$ star, 
orbiting the central object
with a 7.2\,d period and causing the measured 
radial velocity variations with an amplitude of 2.4\,km\,s$^{-1}$, would be 
less massive than 24\,M$_{\mbox{\tiny Jup}}$. 
 
V and B band photometry obtained in 7 consecutive nights in
Chile and Australia shows brightness variations 
of the order of 0.15\,mag in V and 0.2\,mag in B.
The shape of these light curves obtained by us in 2000 differs 
significantly from the shape of the light curves obtained by W98 in 1996.
In particular the deep minima of 0.5\,mag and 0.3\,mag
observed in 1996 are not detected by us. Therefore 
%it is excluded that
the brightness variations recorded in 1996 cannot be explained by 
eclipses of a binary system.

Another argument against the
eclipsing binary hypothesis is the behaviour of the color variations,
since the object
%We recorded (B-V) color variations indicating that the object 
is redder
during minimum light. 
This behaviour is characteristic for (cool and hot) spots, but can
also be caused by the primary eclipse of a binary with a very red, low-mass 
companion. The secondary eclipse of such a system would show the opposite 
color variation, i.e. becoming bluer during minimum light. 
Such a behaviour is not observed during the 7 nights. 

We also checked the V and B band data recorded in 1996 
(W98, B data: pers. comm.) on color variations. 
The emission of the object is shifted during both minima 
towards longer wavelength and therefore none of the minima is
explicable by a 
secondary eclipse of a low-mass companion. 
However, a companion with similar mass as the primary 
would not show color variations at all but this is in any case
definitely excluded by our spectroscopic observations.

Concerning the rotational period of the star we found that our 
photometric data are periodic with about 3.6\,d.
These data are also consistent with the period of 7.2\,d found by
W98. On the other hand the W98 data are also in agreement 
with a 3.6\,d period, apart from small deviations.

We estimated an upper limit for the rotational period
based on the radius and the projected rotational 
velocity. We took into account different distance estimates for the
Lupus star forming region, 
as well as cross checked the radius estimate with different
methods.
In any case
the rotational period of \object{RXJ1608.6-3922} does not exceed
a duration of 6.1 days, 
favoring 3.6\,d as the true rotational period.

The radius was estimated with two different methods:
once from the luminosity and the effective temperature 
and once by means of Barnes-Evans-type relations published
by B99. 
The results agree within the
measurements uncertainties but the B99-radii are 
systematically smaller. 
Depending on the adopted distance the radius is 
1.07\,R$_{\odot}$ (1.50\,R$_{\odot}$, 2.04\,R$_{\odot}$) 
for d=100\,pc (140\,pc, 190\,pc) based on the 
L$_{\mbox{\tiny bol}}$-T$_{\mbox{\tiny eff}}$-estimation and 
0.64\,R$_{\odot}$ (0.89\,R$_{\odot}$, 1.21\,R$_{\odot}$), respectively, 
based on the B99-estimation. 
Since v\,sin\,i for the star is 22.4\,km\,s$^{-1}$, 
a rotational period of 3.6\,d allows the radius not to be
smaller than 1.6\,R$_{\odot}$. 
These considerations suggest a radius
for \object{RXJ1608.6-3922} 
in the range of 1.6 to 2.5\,R$_{\odot}$.
It should be mentioned that 
this result is not compatible with a distance of 100\,pc.

The variations of the photometric brightness of \object{RXJ1608.6-3922}
are interpreted as modulation due to active, spotted regions 
on the stellar surface. They probably also account for the 
observed radial velocity scatter of 2.4\,km\,s$^{-1}$.

Krautter et al. (1997)
classify \object{RXJ1608.6-3922} as a weak line T Tauri star 
on basis of an H$_{\alpha}$ equivalent width of 7.4\,{\AA} 
(spectrum obtained in the time span between 1991 and 1993).
Wichmann et al. (1999) measured an equivalent width of 14.2\,{\AA}
in 1995,
based on high-resolution spectroscopy. These measurements suggest that
the H$\alpha$ emission of this star is variable and that it 
might well be a classical T Tauri star (cTTS). 
The accretion of circumstellar material onto the 
surface of a star can cause dominant hot spots. 
Provided that \object{RXJ1608.6-3922} is a cTTS, the detected variability 
can be caused by such hot spots and regardless of the presence of 
circumstellar material 
the photometric variability can be caused by cool spots. 

The variable shape of the light curve indicates that
the spotted regions change their properties 
(size, temperature and/or location) in time 
scales of a few years.

Variable shapes of light curves are observed for 
many active stars, like RS CVn and BY Dra stars.
For RS CVn stars it is known that their light curves change
in shape, mean light level and amplitude
on time scales of a year or even more rapid 
(e.g. 
V\,711\,Tau (Kang \& Wilson 1989), 
$\sigma$\,Gem (Ol\'ah et al. 1989), 
DM\,UMa (Mohin \& Raveendran 1992), 
HK\,Lac (Ol\'ah et al. 1997)).
This is interpreted as rotational modulation due to cool 
spots on the surface, which rapidly change their location, size and/or 
temperature.

Analysis of the evolution of spots for several pre-main-sequence
stars by Grankin (1998) (and references therein) 
showed that the initial epochs and the 
rotational periods are stable on time scales 
of several years, whereas the shape and amplitude of the 
light curves changes from year to year.   
The stability of the rotational periods are explained
by stable positions of cool, spotted regions confined to
a specific active longitudinal
region. The variations of 
the spot light curve in shape and amplitude
can then be explained by the appearance, disappearance 
and migration of individual spots.
This scenario is also supported by the work of
Sch\"ussler \& Solanki (1992), who 
demonstrate 
that in rapidly rotating stars, 
the strong curiolis forces push the magnetic flux 
always towards the poles. 
Therefore the observations of cool spots remaining at the same position for 
several years on T Tauri stars can be explained by 
stable active regions in which new spots are produced continuously.  

For V\,410\,Tau photometric observations over the time span 
1981 to 1997 (Vrba et al. 1988, Herbst 1989, Grankin 1999) show
that the photometric behaviour of this system is characterized by 
year-to-year variations of the shape and amplitude of the light curve.
Modeling of the spot distribution based on photometry (Herbst 1989),
as well as Doppler imaging of the surface structure
(Hatzes 1995 and references therein)
indicate that 
there is a stable high-latitude spot
area lasting for at least several years.

Another example for a T Tauri star with indications for a long lasting
stable spot distribution is \object{P1724} (Neuh\"auser et al. 1998
and references therein). A periodicity of 5.7\,d is present
in photometric observations covering almost 30 years
and can even be traced in radial velocity measurements
covering about 1.5\,years. 
The relative phasing between the photometric and the spectroscopic 
data is almost stable over the entire time span of the spectroscopic 
observations, suggesting a basically non-varying spot 
distribution.

We conclude that \object{RXJ1608.6-3922} is
a spotted star with a rotational period of
about 3.6 days and that spots have mimicked the light-curve of
an eclipsing binary in the data which were
previously obtained.
There are indications that the stellar activity
on \object{RXJ1608.6-3922} is highly variable,
with an equivalent width of the H$_{\alpha}$
emission changing from about 7 to about 14\,{\AA} within a few years
and a variable shape of the light curve, with amplitudes 
changing from about 0.5 to about 0.2\,mag within 4 years.
Therefore \object{RXJ1608.6-3922} 
is an interesting object for further study of 
stellar activity on T Tauri stars.

\begin{acknowledgements}
      We acknowledge helpful discussions on the topic of this paper with
      G. Wuchterl, N. Hu\'elamo and K. Fuhrmann.
      VJ acknowledges grant from the Deutsche Forschungsgemeinschaft
      (Schwerpunktprogramm `Physics of star formation').
      RN acknowledges financial support from the Bundesministerium
      f\"ur Bildung und Forschung through the Deutsche Zentrum f\"ur
      Luft- und Raumfahrt e.V. (DLR).
\end{acknowledgements}

\end{document}